\documentclass[iop,apj,twocolappendix]{emulateapj}
\usepackage{apjfonts}

\usepackage{graphicx}  
\usepackage{float}
\usepackage{amsmath}
\usepackage{xspace}
\usepackage{aas_macros}
\usepackage{natbib, hyperref}
\usepackage{dcolumn}
\usepackage{amssymb}
\usepackage{ulem}
\usepackage[usenames,dvipsnames]{xcolor}

\hypersetup{
    colorlinks=true, 
    linktoc=all,     
    linkcolor=Blue,  
    citecolor=OliveGreen,
}

\newcommand{\msun}{\mbox{${\rm M}_{\odot}$}\xspace}
\newcommand{\myr}{\mbox {~${\rm M_{\odot}~\rm yr^{-1}}$}}

\newcolumntype{d}{D{.}{.}{2}}

\def\apgt{{\raise-.5ex\hbox{$\buildrel>\over\sim$}}\ }
\def\aplt{{\raise-.5ex\hbox{$\buildrel<\over\sim$}}\ }

\begin{document}

\title{The formation of
  Cataclysmic Variables: the influence of nova eruptions}

\author{G. Nelemans\altaffilmark{1,2}}
\email{nelemans@astro.ru.nl}
\author{L. Siess\altaffilmark{3}}
\author{S. Repetto\altaffilmark{1}}
\author{S. Toonen\altaffilmark{4}}
\author{E.S. Phinney\altaffilmark{5,1}}

\altaffiltext{1}{Department of Astrophysics/IMAPP, Radboud University, Nijmegen, The Netherlands}
\altaffiltext{2}{Institute for Astronomy, KU Leuven, Leuven, Belgium}
\altaffiltext{3}{Institut d'Astronomie et d'Astrophysique, Universite
  libre de Bruxelles (ULB), Brussels, Belgium }
\altaffiltext{4}{Leiden Observatory, Leiden University, Leiden, The Netherlands}
\altaffiltext{5}{Theoretical Astrophysics, 350-17, California Institute of Technology, Pasadena, CA, USA}


\begin{abstract}
The theoretical and observed populations of pre-cataclysmic variables
(pre-CVs) are dominated by systems with low-mass white dwarfs (WDs),
while the WD masses in CVs are typically high. In addition, the space
density of CVs is found to be significantly lower than theoretical
models. We investigate the influence of nova outbursts on the
formation and (initial) evolution of CVs. In particular, we calculate
the stability of the mass transfer in case all the material accreted
on the WD is lost in classical novae, and part of the energy to eject
the material comes from a common-envelope like interaction with the
companion. In addition, we study the effect of an asymmetry in the
mass ejection, that may lead to small eccentricities in the orbit. We
find that a common-envelope like ejection significantly decreases the
stability of the mass transfer, in particular for low-mass
WD. Similarly, the influence of asymmetric mass loss can be important
for short-period systems and even more so for low-mass WD, but likely
disappears long before the next nova outburst due to orbital
circularization. In both cases the mass-transfer rates increase, which
may lead to observable (and perhaps already observed) consequences for
systems that do survive to become CVs. However, a more detailed
investigation of the interaction between nova ejecta and the companion
and the evolution of slightly eccentric CVs is needed before definite
conclusions can be drawn.
\end{abstract}

\maketitle

\section{Introduction}\label{sec:intro}

Cataclysmic Variables \citep[CVs,][]{war95b} have long been recognized
as interacting binaries in which a white dwarf (WD) accretes material
from a companion star via an accretion disk. The systems with
non-magnetic WDs are in the standard picture thought to follow an
evolution that passes through the following stages
\citep[e.g.][]{rvj83,2011ApJS..194...28K}: i) Onset of mass transfer
at periods of several hours. ii) Evolution to shorter periods with
mass transfer driven by magnetic braking leading to mass-transfer
rates of order $10^{-8} \myr$. iii) Cessation of the mass transfer at
periods of about 3 hours leading to a ``period gap''. iv)
Re-establishment of mass transfer at periods of around 2 hours, now
driven by gravitational wave emission, leading to substantially lower
mass-transfer rates of order $10^{-10} \myr$.  v) A period minimum
around 60-80 min, where the mass-transfer rate drops
significantly. A (significant) part of the population may form
  from systems with low-mass donors that join the evolution at stage iv).

In this picture, the progenitors of CVs are wide binaries in which the
intermediate-mass primary evolves to the RGB/AGB stage after which the
substantially lower-mass companion experiences a spiral-in in a
common-envelope phase to end with a WD and a main-sequence (MS) star
in a close binary \cite[e.g.][]{pac76}.

This standard scenario suffers from a number of problems and
shortcomings. One of them is that studies of the potential progenitors
of CVs, find that the majority of progenitors ($\apgt$75\%) have
  low-mass primaries
  \citep[e.g.][]{pw89,dek92,kol93,1996ApJ...465..338P}, leading to a
  predicted mass distribution of WDs in CVs that is dominated by
  low-mass WDs, contrary to the observed trend that WDs in CVs are
  massive \citep[$M_{\rm WD} > 0.7 \msun$, see][and references
    therein]{2011A&A...536A..42Z}. In addition, the theoretical models
  predict a rather large space density of CVs, compared to
  observational estimates \citep[see][and references
    therein]{2014xru..confE.164P}. There could be a number of
  different solutions to these problems. On the one hand, it could be
  that the observational estimates are dominated by selection effects:
  there exist in fact many CVs and most of them have lower-mass WDs,
  but we predominantly see the small number of systems that have
  massive WD. However, recent much more homogeneous samples of CVs, in
  particular those found in SDSS, make this argument not very
  convincing \citep[see][]{2011A&A...536A..42Z,2009MNRAS.397.2170G}.
  A second solution could be that for some reason the true space
  density of CVs is lower than in the models and that the WDs in CVs
  actually grow in mass so that the massive WDs we see now were in
  fact low-mass WDs when the CVs formed
  \citep[e.g.][]{2014MNRAS.441..354T}. This option is studied in
  detail by \citet{2015A&A...577A.143W}, who conclude that it is
  unable to explain the observed WD mass distribution. A third
  solution could be that for some reason the theoretical models are
  incomplete and, for instance, the common-envelope phase through
  which the CV progenitors go \citep{pac76} preferentially selects
  massive WDs to become CVs. Yet, this hypothesis is ruled out because
  the direct progenitors of CVs (WD with MS companions) are observed
  and do preferentially have relatively low-mass WDs
  \citep{2011A&A...536A..42Z}.  In this paper we study the
  alternative, that in fact the lower-mass WD that exist in the pre-CV
  population do not make it to become long-lived CVs because they
  merge due to additional angular momentum loss or induced small
  eccentricity in the first nova
  outbursts. \citet{2015arXiv151004294S} independently came to the
  conclusion that the lower-mass WDs could be removed from the sample
  if for some reason they suffer extra angular momentum loss. In
  Sect.~\ref{stability} we review the factors that determine the
  stability of the mass transfer. In Sect.~\ref{novae} we derive ways
  to estimate the effect of classical nova outburst on the mass
  transfer stability, first for a brief common-envelope phase
  (Sect.~\ref{CE}), then for rapid asymmetric mass loss
  (Sect.~\ref{ecc}). In Sect.~\ref{results} we show the results of our
  calculations for different assumptions. In Sect.~\ref{discussion} we
  discuss how our findings fit in the theoretical and observational
  knowledge about novae. In Sect.~\ref{conclusion} we summarize our
  conclusions.

\section{The stability of mass transfer}\label{stability}

When the MS star in a WD -- MS binary first fills its Roche lobe, a
complex process starts, in which material is transferred from
the MS star to the WD, changing the mass ratio of the system, which
in turns changes the orbital separation (and thus the size of the
Roche lobe). In addition, (some) of the material may not end up on the
WD, but leave the binary, taking away angular momentum. Finally, the
MS star will change its radius owing to its loss of mass. The net
effect will be a change in the relative size of the MS star radius to
its Roche lobe, which drives the mass transfer to either go up, go
down or stay the same. Because the radius change of the MS star
depends on the speed at which mass is lost, the final result
can be
\begin{itemize}
\item stabilization of the mass transfer on a time scale such that the
  MS is roughly in equilibrium. The mass-transfer rate is set by the
  time scale of the angular momentum loss from the binary, through
  magnetic braking, gravitational wave radiation and mass loss from
  the system.
\item stabilization of the mass transfer on a shorter time scale. The
  MS star tries to evolve back to thermal equilibrium on
  its thermal time scale and this is the time scale that sets the
  mass-transfer rate \citep[e.g.][]{2002ASPC..261..242S}.
\item the mass transfer does not stabilize and the system most likely
  merges to become a single object, consisting of the WD surrounded
  by the mass of the MS star.
\end{itemize}

\begin{figure}
\centering
\includegraphics[width=\columnwidth]{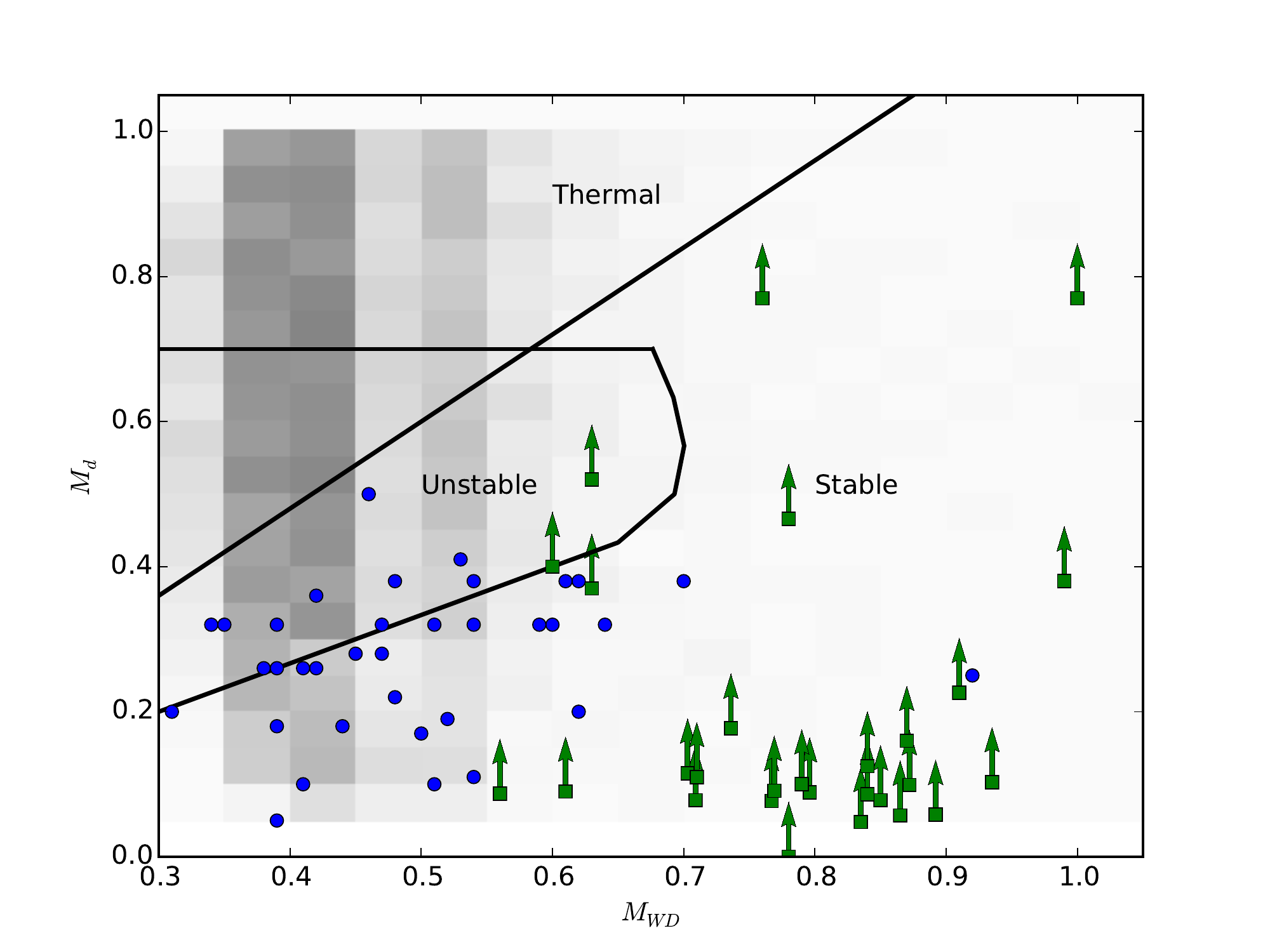}
\caption{Accretor versus donor mass at the onset of mass transfer. The
  lines indicate the theoretical stability limits, below which the
  mass transfer is expected to be stable. The grey shade shows the
  distribution of these parameters for the best model of post
  common-envelope binaries from \citet{2013A&A...557A..87T}. The blue
  circles show the pre-CV systems, while the green squares with arrows
  indicate the masses in CVs (both from \citet{2011A&A...536A..42Z})}
\label{fig:Mwd_Mdonor}
\end{figure}

In Fig.~\ref{fig:Mwd_Mdonor} we show the expected stability regions
for WD -- MS stars when filling their Roche lobe, assuming
conservative mass transfer (all material lost from the donor is
accreted by the WD). The regions are taken from \citet{pw89}, and are
based on two limits: for low-mass MS stars (below 0.7 \msun, that have
a significant convective envelope), the mass transfer is expected to
be unstable if the mass ratio ($M_{\rm donor}/M_{\rm WD}$) is larger
than 2/3 (marked ``Unstable'' in the figure). For MS masses approaching
the 0.7 \msun this limit becomes larger (smoothly curving to a mass
ratio around one). For MS stars above 0.7 \msun, with mainly radiative
envelopes, the mass transfer is expected to be stable for large mass
ratio's, but above 1.2 would proceed on the thermal time scale (marked
``Thermal'' in the figure, above the diagonal line).

In the same figure we show the known pre-CVs and the CVs with known WD
masses \citep[both from][]{2011A&A...536A..42Z}, where the arrows
indicate that in the CVs the donors could have started mass transfer
at higher mass. In grey we plot a theoretical pre-CV
population, showing the the preference for low-mass WD. It is
  model $\alpha\alpha2$ described in \citet{2013A&A...557A..87T},
  which was found to fit best with the observed post common-envelope
  binary population.

\section{Classical novae and their influence on the evolution}\label{novae}

The above theoretical stability limits are based on overly simplified
assumptions, in particular that all the transferred mass stays on the
WD. Accreting WD accumulate the accreted material in a layer and when
the density and temperature at the bottom of the layer are high
enough, nuclear fusion in the layer causes a classical nova outburst
\citep{1972ApJ...176..169S,tb04}, in which much, if not all, accreted
mass is lost from the system. This causes different effects on the
binary evolution: the mass loss can widen the binary, lowering the
mass-transfer rate. On the other hand, if the expanding envelope
interacts strongly with the companion, the ejected mass could take
along relatively large amounts of angular momentum, shrinking the
orbit \citep[e.g.][]{1992ASPC...22..316L}. Finally, if the mass-loss
happens fast and is asymmetric, it may induce a small eccentricity in
the orbit that may influence the mass
transfer. \citet{1986ApJ...311..163S} have studied the influence of
novae on the orbit and concluded that in principle CVs could have long
periods of ``hibernation'', in which the binary becomes detached and
mass transfer ceases. This happens if the nova outburst ejects the
mass rapidly without much scope for interaction with the
companion. However, recent observations of nova outbursts suggest that
the ejecta are in fact strongly influenced by the companion
\citep[e.g.][]{2009ApJ...706..738W,2013ApJ...768...49R,2014Natur.514..339C}. We
therefore below study how such interactions could affect the stability
of the mass transfer.

\subsection{Angular momentum loss in a common-envelope}\label{CE}

For a given formalism that describes the change in orbital separation
due to a common-envelope-like phase we can determine the associated
angular momentum loss. This then can be added to the other angular
momentum losses to calculate the stability of the mass transfer
\citep[e.g.][]{1991A&A...246...84L,2015ApJ...805L...6S}.

We assume here that the nova eruption leads to expansion of the
envelope and that at the time this envelope reaches the companion star
(i.e. at a radius equal to the orbital separation) the friction of the
common-envelope takes over the energy generation to bring the
material to infinity. Of course the nuclear burning in principle can
provide enough energy to eject the envelope (if it is not radiated) so
we simply assume the common-envelope's orbital energy is used to eject
a fraction $f_{\rm CE}$ of the material, the rest being ejected by the
energy from the burning. To calculate the angular momentum loss
associated with the common-envelope, we here consider only this
fraction of the ejected mass $M_{\rm ej} = f_{\rm CE} M_{\rm
  accreted}$ and can write its binding energy as
\begin{equation}
E_{\rm bind} = \frac{G M_{\rm WD} M_{\rm ej}}{a_i},
\end{equation}
while the orbital energy is given by
\begin{equation}
E_{\rm orb, i} = \frac{G M_{\rm WD} M_{\rm d}}{2 a_i}.
\end{equation}
The final orbital energy then is
\begin{equation}\label{eq:Ef}
E_{\rm orb, f} = E_{\rm orb, i} -  E_{\rm bind} =
\frac{G (M_{\rm WD} - M_{\rm ej}) M_{\rm d}}{2 a_f}.
\end{equation}
Rearranging the terms and writing out the last term of eq.~\ref{eq:Ef} we
get
\begin{equation}
\frac{2 M_{\rm WD} M_{\rm ej} + M_{\rm WD} M_{\rm d}}{2 a_i} =
\frac{M_{\rm WD} M_{\rm d} - M_{\rm ej} M_{\rm d}}{2 a_f}
\end{equation}
so
\begin{equation}
\frac{a_f}{a_i} = \frac{1 - M_{\rm ej}/M_{\rm WD}}{1 + 2
  M_{\rm ej}/M_{\rm d}}
\end{equation}
which (because $M_{\rm ej} \ll M_{\rm WD}, M_{\rm d}$) is well
approximated by
\begin{equation}
\frac{a_f}{a_i} \approx 1 - M_{\rm ej}/M_{\rm WD} - 2  M_{\rm ej}/M_{\rm d}
\end{equation}
so with $q = M_{\rm d}/M_{\rm WD}$, the relative change in orbit is
\begin{equation}
\frac{\Delta a}{a} = \frac{a_f - a_i}{a_i} = -\frac{M_{\rm
    ej}}{M_{\rm d}} \left(2 + q\right).
\end{equation}
This result is somewhat different (smaller) than eq. (2) of
\citet{2015ApJ...805L...6S}, who considers as binding energy the
energy needed to bring the envelope to infinity from the L1 point.

The change in orbital angular momentum (due to the change in
separation and mass) is
\begin{equation}
\frac{\Delta J_{\rm orb}}{J_{\rm orb}} = -\frac{M_{\rm
    ej}}{M_{\rm d}} \left(1 + q + \frac{q^2}{2 (1 +
  q)}\right).
\end{equation}
 For values of $q$ between 0.2 and 1 (relevant for CV systems) the
 above expression is within ten per cent of the simple and often used
 angular momentum loss from a ``circumbinary'' ring with radius of $a$
 \citep{sph97,2006csxs.book..623T}.
\begin{equation}\label{eq:J_CE}
\frac{\Delta J_{\rm orb}}{J_{\rm orb}} = -\frac{M_{\rm
    ej}}{M_{\rm d}} (1 + q)
\end{equation}

We performed MESA
\citep[][rev. 7184]{2013ApJS..208....4P,2015ApJS..220...15P}
calculations of the evolution of CVs for several different assumptions
for the angular momentum loss due to nova eruptions. For a grid of
donor masses and accretor masses we used the standard magnetic braking
prescription of MESA \citep[based on][]{rvj83} to simulate the
evolution from an orbital period slightly longer than the one at which
Roche-lobe overflow starts. We only simulate the donor star in detail
and prescribe the mass and angular momentum loss from the system as a
combination of isotropic re-emission
\citep[see][]{sph97,2006csxs.book..623T} and mass and angular momentum
loss due to a common-envelope-like process according to
eq.~\ref{eq:J_CE}. We can model the latter as a continuous process,
because the recurrence time between the novae is significantly shorter
than any of the relevant time scales of the donors star, so the MESA
calculations actually use time steps longer than the recurrence
time. We classify the mass transfer as unstable, if it reaches above
$\sim 10^{-4}$ \myr when it is at least a factor of 1000 larger than
thermal time scale mass-transfer and the code breaks down.

\subsection{White dwarf kicks due to asymmetric mass loss}\label{ecc}

\begin{figure*}
\centering
\includegraphics[width=\columnwidth]{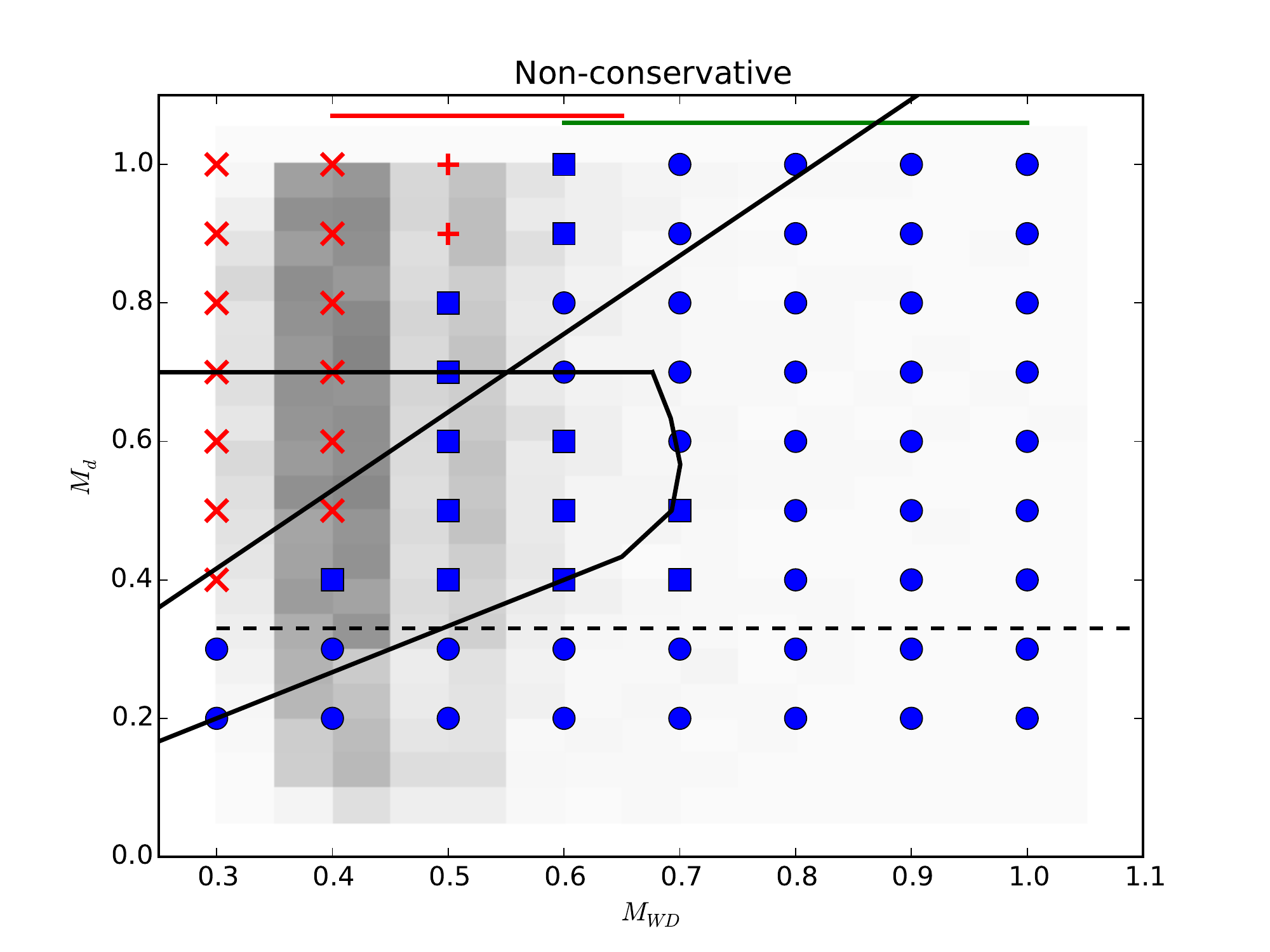}
\includegraphics[width=\columnwidth]{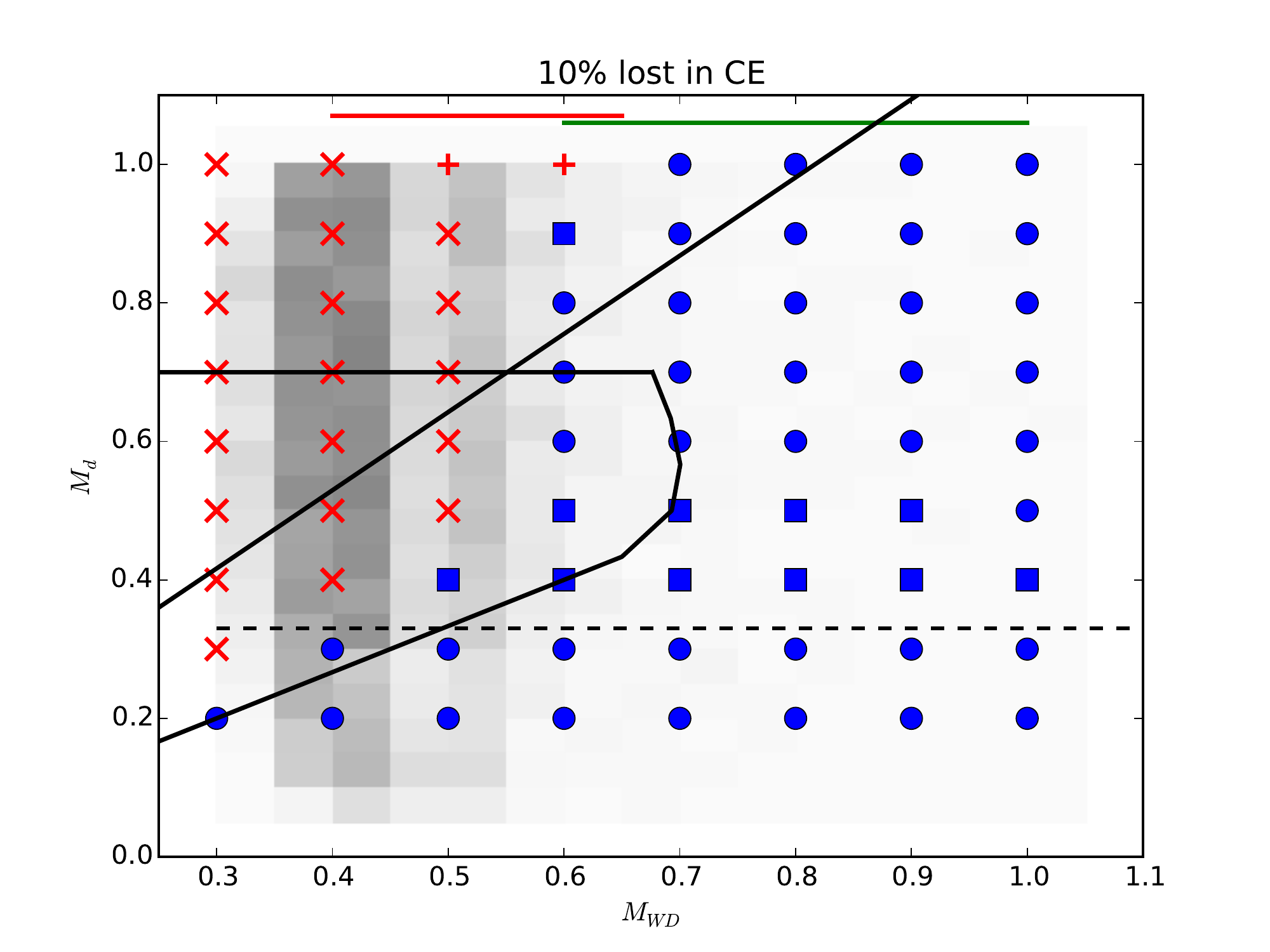}
\includegraphics[width=\columnwidth]{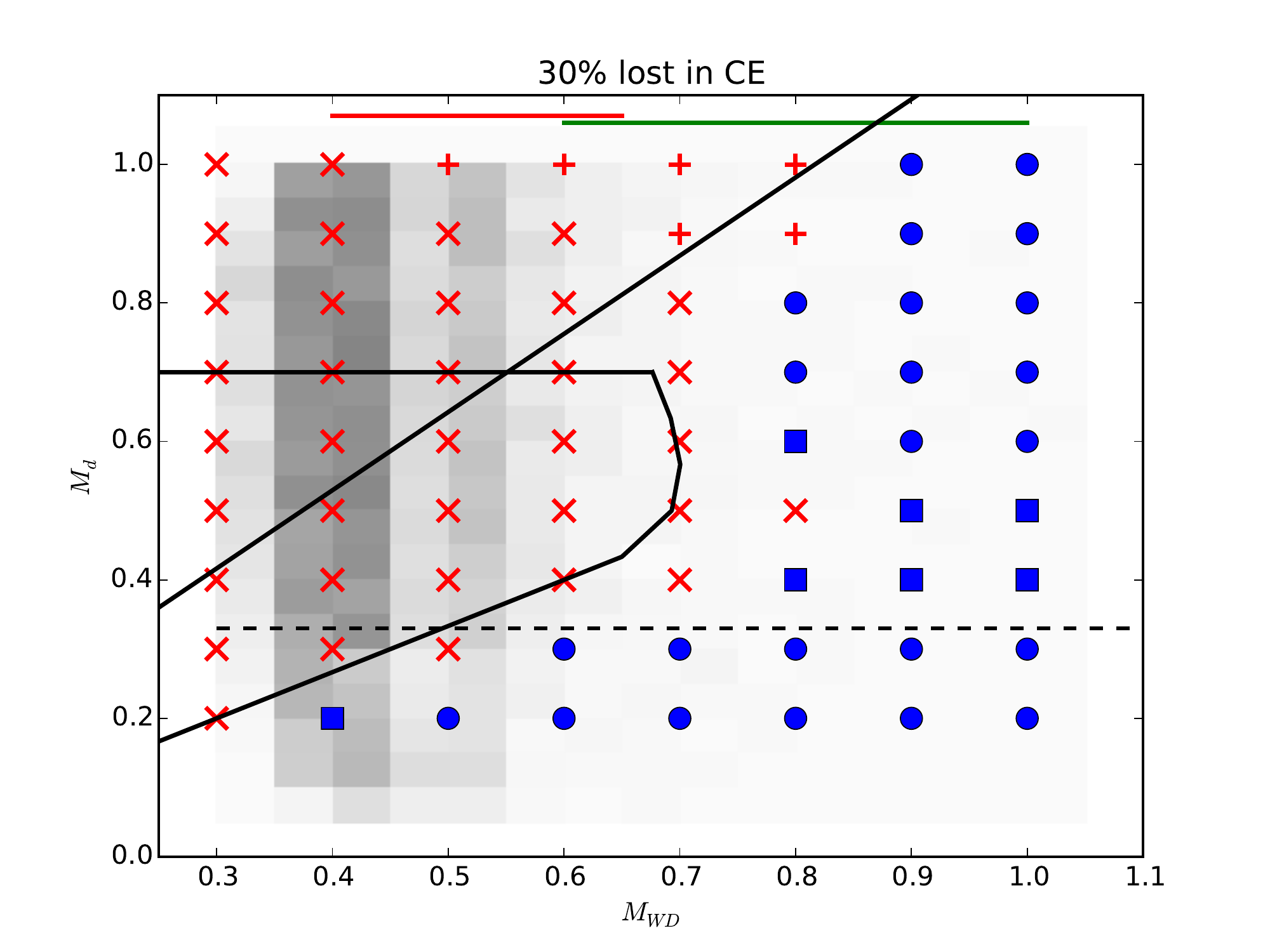}
\includegraphics[width=\columnwidth]{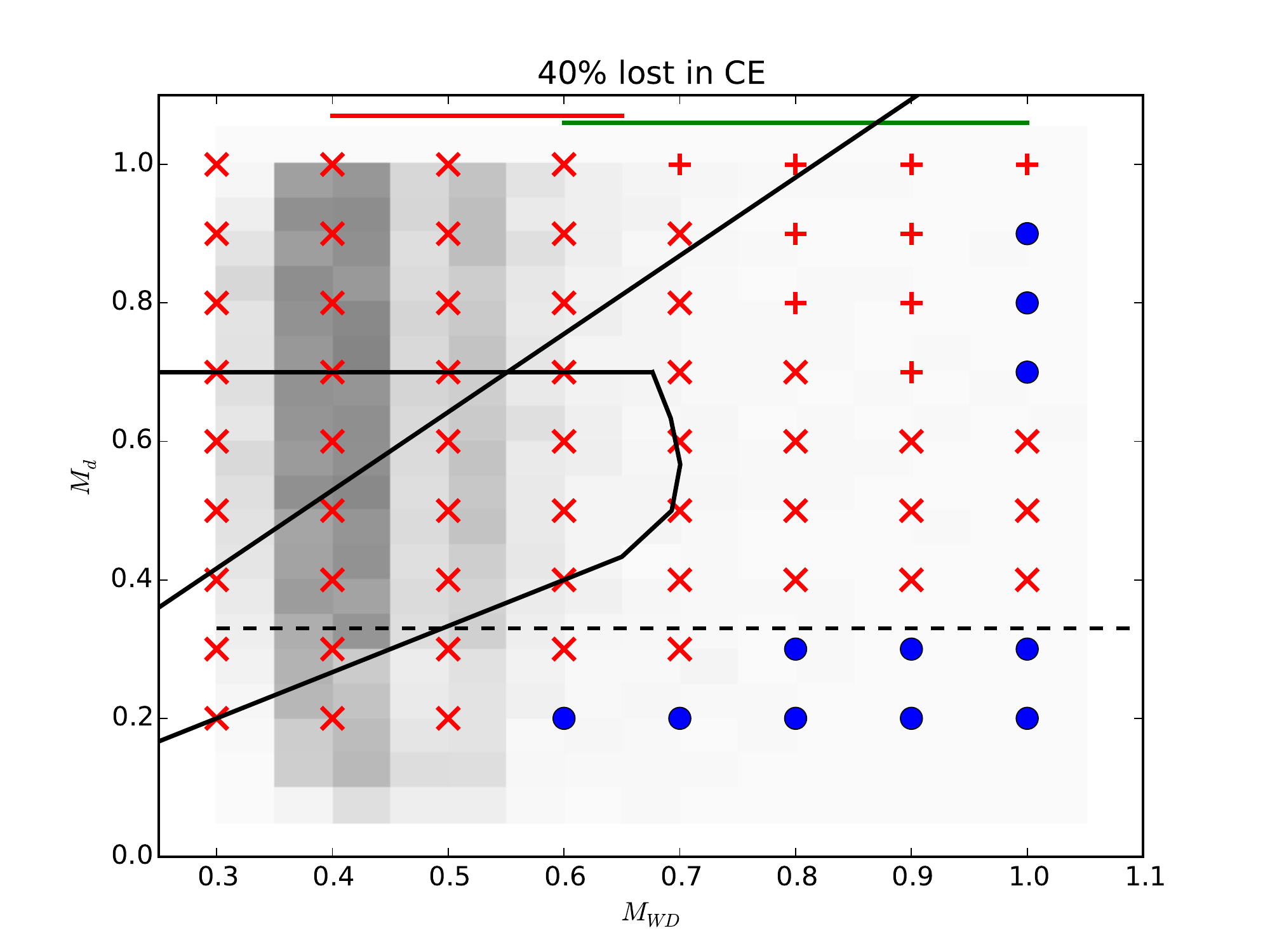}
\caption{Grid of initial accretor versus donor mass for the MESA
  calculations at the onset of mass transfer. The lines and grey shade
  denote the stability limits and theoretical population as in
  Fig.~\ref{fig:Mwd_Mdonor}. The symbols indicate the outcome of the
  MESA calculations. (red) cross: directly unstable, (red) plus:
  unstable after a brief stable phase; (blue) square: thermal time
  scale stable, (blue) circle: stable. The dashed lines give the
  separate onset of mass transfer above and below the period gap. The
  different plots are for for $f_{\rm CE} = 0.0$, i.e. fully
  non-conservative mass transfer, $f_{\rm CE} = 0.1, 0.3$ and 0.4. }
\label{fig:stability0.0}
\end{figure*}

Alternatively, if (part of) the envelope is ejected asymmetrically in
a fast nova eruption, the accreting white dwarf will get a small
velocity kick to conserve linear momentum. As in the case of an
asymmetric supernova explosion that gives a kick to newly formed
neutron stars, this kick will introduce an eccentricity in the
orbit. We performed a Monte Carlo calculation of the effect of a
small, isotropic kick on the orbit using the same method as in
\citet{2015MNRAS.453.3341R} and found that depending on the direction
of the kick, the semi-major axis either increases or decreases, but
that in the vast majority of the case the periastron distance in the
new orbit is smaller than the pre-nova separation. This could lead to
a (strong) increase in the mass transfer rate at periastron. In order
to estimate the maximum effect of asymmetric mass loss, we calculate
the most extreme case, in which the kick is directed opposite to the
orbital velocity of the WD.  We assume the mass is leaving the
accreting white dwarf with an ejection velocity $v_{\rm ej}$. The
resulting kick velocity of the white dwarf $v_{\rm kick}$ is given by
\begin{equation}
v_{\rm kick} = f_{\rm kick} v_{\rm ej}  \frac{M_{\rm ej}}{M_{\rm
    WD}}
\end{equation}
where $f_{\rm kick}$ is the fraction of the mass that is ejected
asymmetrically. With $v_{\rm ej} = 500 - 3000$ km/s \citep[][and
  references therein]{2012BASI...40..267C,2013ApJ...768...49R}, and
$M_{\rm ej}/M_{\rm WD} \aplt 10^{-3}$ the kick could be up to a km/s.

For initial orbital separation $a_0$, the eccentricity and semi-major
axis after the kick can be derived in the relevant small-change limit
showing its main effects. For our actual calculations below we use the
full equations \citep[e.g.][]{bp95,1996ApJ...471..352K}. Defining
\begin{equation}
M_{\rm tot,f}/M_{\rm tot,i}=1-\delta
\end{equation}
(so $\delta > 0$ is the fractional change in total mass), and
\begin{equation}
\nu=v_{\rm kick}/v_{\rm rel}
\end{equation}
where $v_{\rm rel}$ is the relative velocity of the two stars (and
taking $\nu > 0$ when the kick is directly opposed to $v_{rel}$), one
gets\footnote{An interesting point is that the pericenter is then
  given, to linear order in $\delta$ and $\nu$, by $r_p = 1 - 2(2\nu -
  \delta)$ for $2\nu > \delta$; $r_p = 1$ for $2\nu < \delta$,
  i.e. unless the kick velocity is greater than $(1/2)\delta v_{\rm rel}$,
  about 10m/s for a typical case, the initial semi major axis is the
  pericenter, not the apocenter, so no enhanced Roche lobe overflow is
  possible. }
\begin{equation}
\frac{a_f}{a_i} = 1-2\nu+\delta   
\end{equation}
and the resulting eccentricity is
\begin{equation}
e_{\rm strong} = |2\nu - \delta|
\end{equation}

To estimate the effect on the mass-transfer rate we calculate the
Roche-lobe overfill factor $\Delta = (R_* - R_L)$ as function of the
orbital phase ($\phi$), assuming to first order that the relative
change in the Roche lobe follows the relative change in the separation
and assuming that before the nova $\Delta = 0$. For the relevant case
$2\nu > \delta$,
\begin{equation}
\Delta(\phi)/R_\ast = -(1+q)\delta\frac{\partial\ln R_L}{\partial\ln q}
                       + (2\nu-\delta)(1+\cos\phi)
\end{equation}
with $\frac{\partial\ln R_L}{\partial\ln q}$ derived from the
Roche-lobe approximation, e.g. \citet{egg83}.  For small values of
$\Delta$ the mass-transfer rate scales as \citep{1988A&A...202...93R}
\begin{equation}\label{eq:Mdot}
\dot{M} \propto e^{\Delta/H},
\end{equation}
with 
\begin{equation}
H = \frac{k_B T_{\rm eff}}{\mu m_H g},
\end{equation}
the pressure scale height of the MS atmosphere.

In order to determine the eccentricities that could arise from
asymmetric mass loss, we have to calculate the effects of single
novae, and therefore have to assume ignition masses and mass transfer
rates. We take the ignition masses from \citet{tb04}, assuming mass
transfer rates of $10^{-8}-10^{-9} \myr$ above the period gap and
$10^{-10} \myr$ below the period gap. We then calculate the effect of
kick on the orbit and the mass-transfer rate, find the new ignition
mass and calculate the time to the next nova which we compare to the
circularization time scale \citep[taken from][]{vp95}.

Furthermore, we use the BINSTAR code described in
\citet{2013A&A...550A.100S,2013A&A...556A...4D}, that performs mass transfer calculations
in eccentric orbits with a full stellar evolution code, to test the
above simplified treatment.

\section{Results}\label{results}

\subsection{Angular momentum loss in a common-envelope like phase}

We calculated the stability of mass transfer for a grid of initial WD
and MS stars for different values of $f_{\rm CE}$, assuming the rest
of the material is lost in a fast symmetric ejection.  In
Fig.~\ref{fig:stability0.0} we show the results for $f_{\rm CE} =
0,0.1,0.2, 0.4$. The fully non-conservative case (top left), with no
common-envelope interaction, shows that in that case significantly
more systems are stable than for the theoretical conservative
limits. A large fraction of the pre-CVs with low-mass WDs would evolve
into CVs, dominating the population both above and below the period
gap. Increasing the fraction of mass ejected via a common-envelope
like process strongly reduces the number of stable systems, in
particular for low-mass WD. For $f_{\rm CE} = 0.1$ the results come
close to the theoretical conservative boundaries, while for $f_{\rm
  CE} = 0.3$ they become more constraining. In both cases the
additional angular momentum loss causes systems that start Roche-lobe
overflow just above the period gap to briefly experience a (very)
short phase of thermal-time scale mass transfer before settling down
on the magnetic braking time scale.  For $f_{\rm CE} = 0.4$ also a
significant fraction of the pre-CVs with massive WDs become unstable
and virtually only systems that start mass transfer below the period
gap remain stable.

\subsection{Eccentric orbits due to asymmetric mass loss}

\begin{figure}
\centering
\includegraphics[width=\columnwidth]{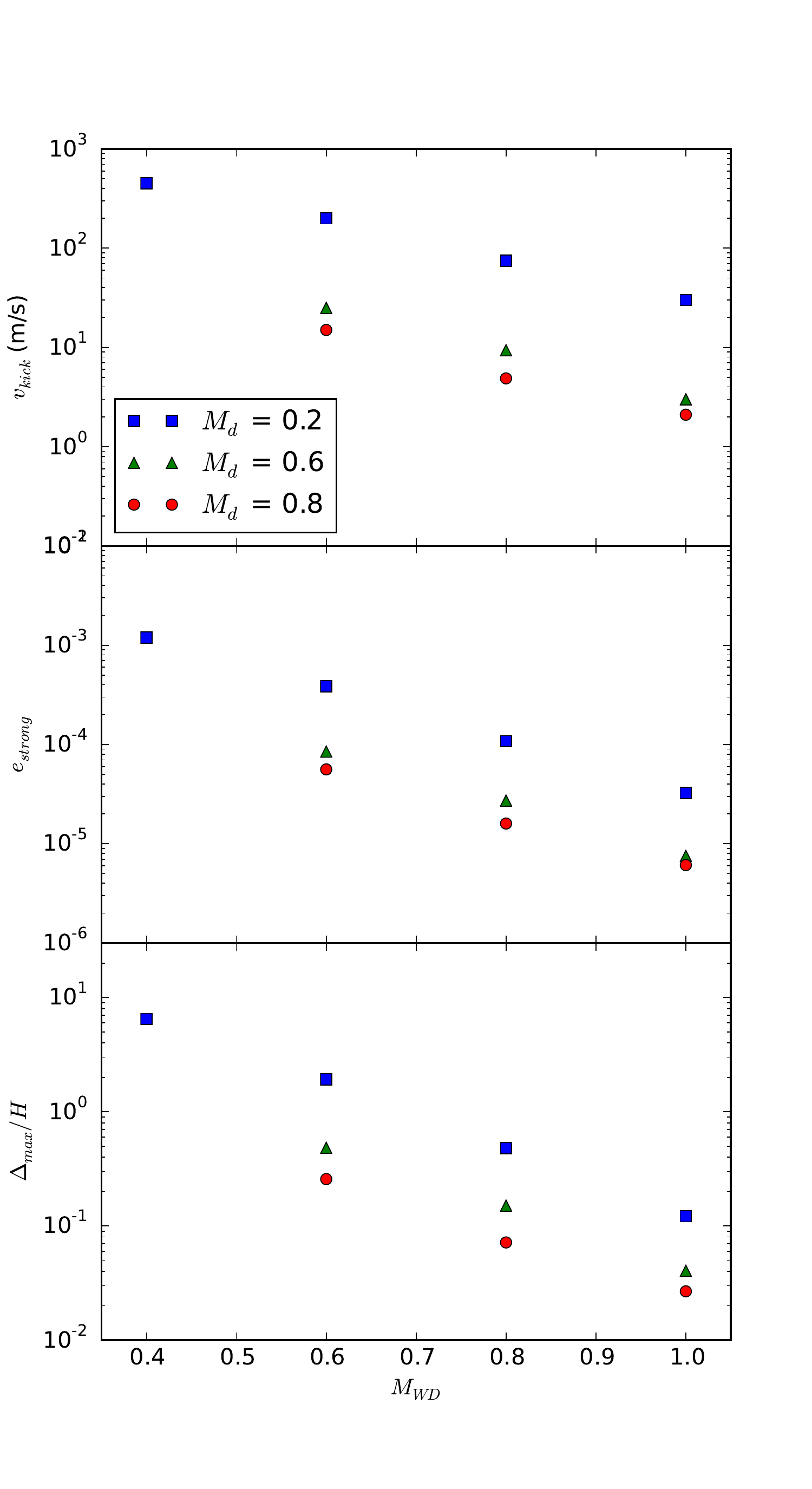}
\caption{Kick velocities, eccentricities and maximum change in
  Roche-lobe overfill factor for novae with $f_{\rm kick} = 0.1$}
\label{fig:v_kick}
\end{figure}

For a more sparse set of initial binaries we calculate the kick
velocity, eccentricity and effect on the mass transfer rate for an
assumed ejecta velocity of 1500 km/s and assuming an asymmetric mass
fraction of 20 per cent, $f_{\rm kick} = 0.2$. The masses and ignition
masses (taken from fig. 9 of \citealt{tb04}) we use are shown in the
first three columns of Table~\ref{tab:results}, the resulting
eccentricity ($e_{\rm strong}$), kick velocity and maximum change in
the Roche-lobe overfill factor ($\Delta$) compared to the donor's
pressure scale height in the next three columns. The latter three are
graphically shown in Fig.~\ref{fig:v_kick}.  We assume $M_{\rm ej} =
M_{\rm ign}$.

\begin{table}
\caption{Resulting $e_{\rm strong}, \Delta$ and global increase in
  mass-transfer rate ($f_{\dot{M}}$) for the different systems
    considered. For the systems with low-mass donors, where the effect
    can be significant, we also calculate the ignition mass for the
    increased mass-transfer rate, its recurrence time and the tidal
    circularization time scale.}
\label{tab:results}
\begin{tabular}{llllllllll}
$M_{\rm WD}$ & $M_{\rm d}$ & $\frac{M_{\rm ign}}{10^{-5}}$ & $e_{\rm strong}$ & $v_{\rm
  kick}$ & $\frac{\Delta_{\rm max}}{H}$ & $f_{\dot{M}}$ &  $\frac{M'_{\rm ign}}{10^{-5}}$&
  $t_{\rm rec}$ & $\tau_{\rm tide}$ \\
(\msun) & (\msun) & (\msun) & ($10^{-4}$) & (m/s) &  & & (\msun) &
  (yr) & (yr)  \\ \hline
1.0 & 0.2 & 10 & 0.3 & 30 & 0.12 & 1.0 & 10 &  1.e6 & 185\\
0.8 & 0.2 & 20 & 1.1 & 75 & 0.5 & 1.2 & 20 & 1.7e6 & 171\\
0.6 & 0.2 & 40 & 3.9 & 200 & 1.9 & 2.9 & 20  & 6.7e5 &  157 \\
0.4 & 0.2 & 60 & 12 & 450 & 6.5 & 144 & 10/S?\footnote{steady burning} & 6.9e3 & 146 \\
1.0 & 0.6 & 1 & 0.08 & 3.0 & 0.04 & 1.0 &  & & \\
0.8 & 0.6 & 2.5 & 0.27 & 9.4 & 0.215& 1.1 &  & & \\
0.6 & 0.6 & 5. & 0.85 & 25 & 0.5 & 1.3 &  & & \\
1.0 & 0.8 & 0.7 & 0.06 & 2.1 & 0.03 & 1.0 &  & & \\
0.8 & 0.8 & 1.3 & 0.16 & 4.9 & 0.07 & 1.0 &  & & \\ 
0.6 & 0.8 & 3.0 & 0.56 & 15 & 0.26 & 1.1 &  & & \\ \hline
\end{tabular}
\end{table}

For the massive donors, i.e. systems above the period gap, the
resulting kicks are typically very small, of order of several m/s,
resulting in very small eccentricities. The change in the overfill
factor then is only a fraction of the scale height and very little
change in the system is expected. For the systems with a 0.2 \msun
donor, the kicks are higher, reaching 500 m/s for the lowest mass
WDs. For these systems the eccentricity reaches $10^{-3}$ and the
orbits change so much that the overfill factor changes by several
scale heights.

We numerically integrate the average increase of the mass-transfer
rate over one orbit compared to the pre-nova circular orbit, using
eq.~(\ref{eq:Mdot}) and show the results in the next column in
Table~\ref{tab:results}. As before, for the systems above the period
gap there is hardly any change. However, for the short period systems
there is a significant change. For the 0.4 + 0.2 \msun system, the
average mass transfer rate is expected to increase by a factor larger
than 100. In order to estimate the effect on the system, we look up
the appropriate ignition masses for these new mass-transfer rates in
\citet{2005ApJ...628..395T} and calculate the time it would take the
system to experience another nova (columns 8 and 9). They are
significantly shorter than the millions of years in unperturbed
systems, but still much longer than the tidal circularization time
scales for the binaries that we calculate using eq.~(2) of
\citet{vp95}, which for these very close binaries are only of order of
100 years. So unless the enhanced mass-transfer rate leads directly to
mass loss from the system (e.g. through the L2/L3 points) that could
influence the further evolution, the effect of asymmetric mass loss
seems short lived, providing only a relatively small increase in the
average mass-transfer rate between novae. For the most extreme
system, the mass-transfer rates increase so dramatically, that the
system may actually get into the regime where the newly accreted
material is burnt directly and stably to helium, rather than
accumulated \citep[indicated by ``S?'' in the table, see fig. 1
  of][]{2005ApJ...628..395T} and the system might show up (briefly) as
a super-soft X-ray source \citep[see][]{hbn+92}.

\subsection{The influence of eccentricity on the evolution}

\begin{figure}
\centering
\includegraphics[width=\columnwidth]{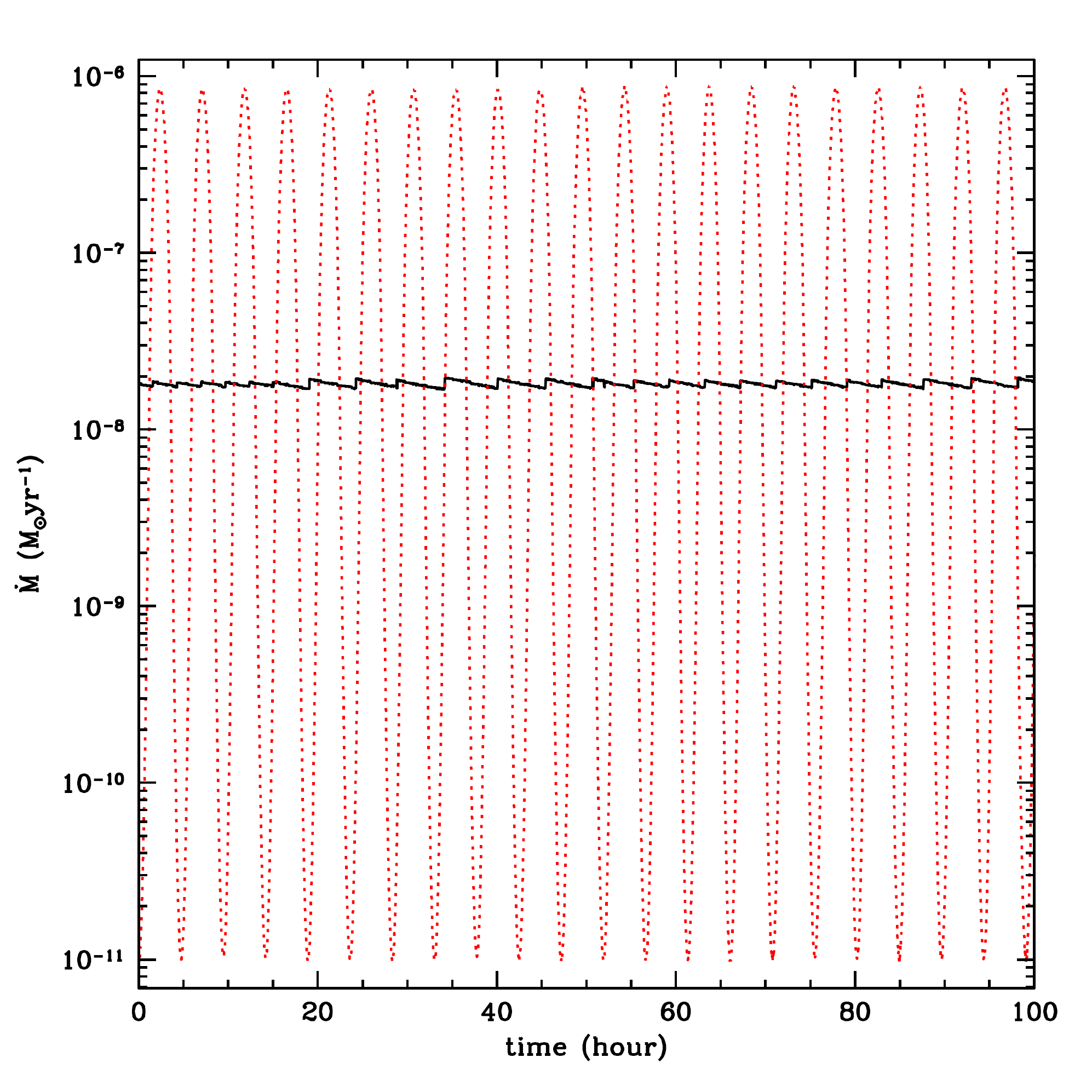}
\caption{Evolution of the mass transfer rate in a 0.6 + 0.6 \msun,
  slightly eccentric ($e= 2 \times 10^{-3}$) orbit in red, compared to
  a circular orbit in black.}
\label{fig:mdot_time}
\end{figure}

 As a test case we evolved a 0.6 white dwarf + 0.6 main sequence star
 with a relatively large eccentricity $e = 2 \times 10^{-3}$ using the
 BINSTAR code (Fig.~\ref{fig:mdot_time}). We started the system in
 such an orbit, that the semi-major axis is equal to the pre-nova
 orbital separation. The mass-transfer rate thus alternating increases
 and decreases compared to the pre-nova mass-transfer rate, for which
 we use $2 \times 10^{-8} \myr$.  The mass transfer in the eccentric
 case indeed varies strongly, with the maximum almost a factor 50
 higher than the pre-nova rate. On average the mass-transfer rate is
 more than a factor 10 higher than in the circular case. For
 comparison, for these parameters, our simple calculation as in
 Sect.~\ref{ecc} gives a factor 100, i.e. overestimates the effect. It
 is clear, that in order to fully assess the influence of such small
 asymmetric mass loss, a systematic study including all the different
 effect should be undertaken, which is beyond the scope of this paper.

\section{Discussion}\label{discussion}

The results show that potentially a common-envelope like phase and
asymmetric mass loss can significantly change the evolution of
CVs. The two main questions are if these effects actually happen and
if so, if they change the stability of the systems in such a way that
the discrepancies between the theoretical and observed CV population
disappear.

From Fig.~\ref{fig:stability0.0} it is clear that for the mechanism to
work comfortably, the systems with low-mass WD should eject a fairly
significant fraction ($\sim$40 per cent) of the mass via a
common-envelope like mechanism, while more massive WDs should be
affected less in order to avoid a deficit of systems above the period
gap. There is no a priori reason to assume the fraction would be the
same. The ejecta velocities are expected to be lower and envelope
masses higher for lower-mass WD, which could lead to more and stronger
interaction of the envelope with the companion
\citep[see][]{1991A&A...246...84L}. Indeed,
\citet{2009ApJ...699.1293K,2011ApJ...743..157K} find that optically
thick winds that drive the mass loss always happen on WD with mass
above 0.7 \msun, but not below, where instead a static giant-like
envelope is found initially. They suggest that for lower-mass WD a
common-envelope like interaction may trigger the transition to a
(wind) mass losing structure. On the other hand, for asymmetric mass
loss to produce a kick, the ejection should happen on a short time
scale compared to the orbital period, so would most likely be
diminished if the nova was slow.

\citet{2015arXiv151004294S} find that a parametrized angular momentum
loss, where the specific angular momentum loss is inversely
proportional to the WD mass works well in an analytic model for the
stability of the mass transfer. The WD mass distribution of the
resulting CV population shows a very good agreement with the observed
WD mass distribution.  \citet{2013IAUS..281...88W} suggests that the
''transient heavy element absorbing'' gas seen in many nova spectra is
due to significant mass loss from the disks in the system, most likely
to a circumbinary disk, which would lead to additional angular
momentum loss with the same scaling as our eq.~(\ref{eq:J_CE}).

\begin{figure}
\centering
\includegraphics[width=\columnwidth]{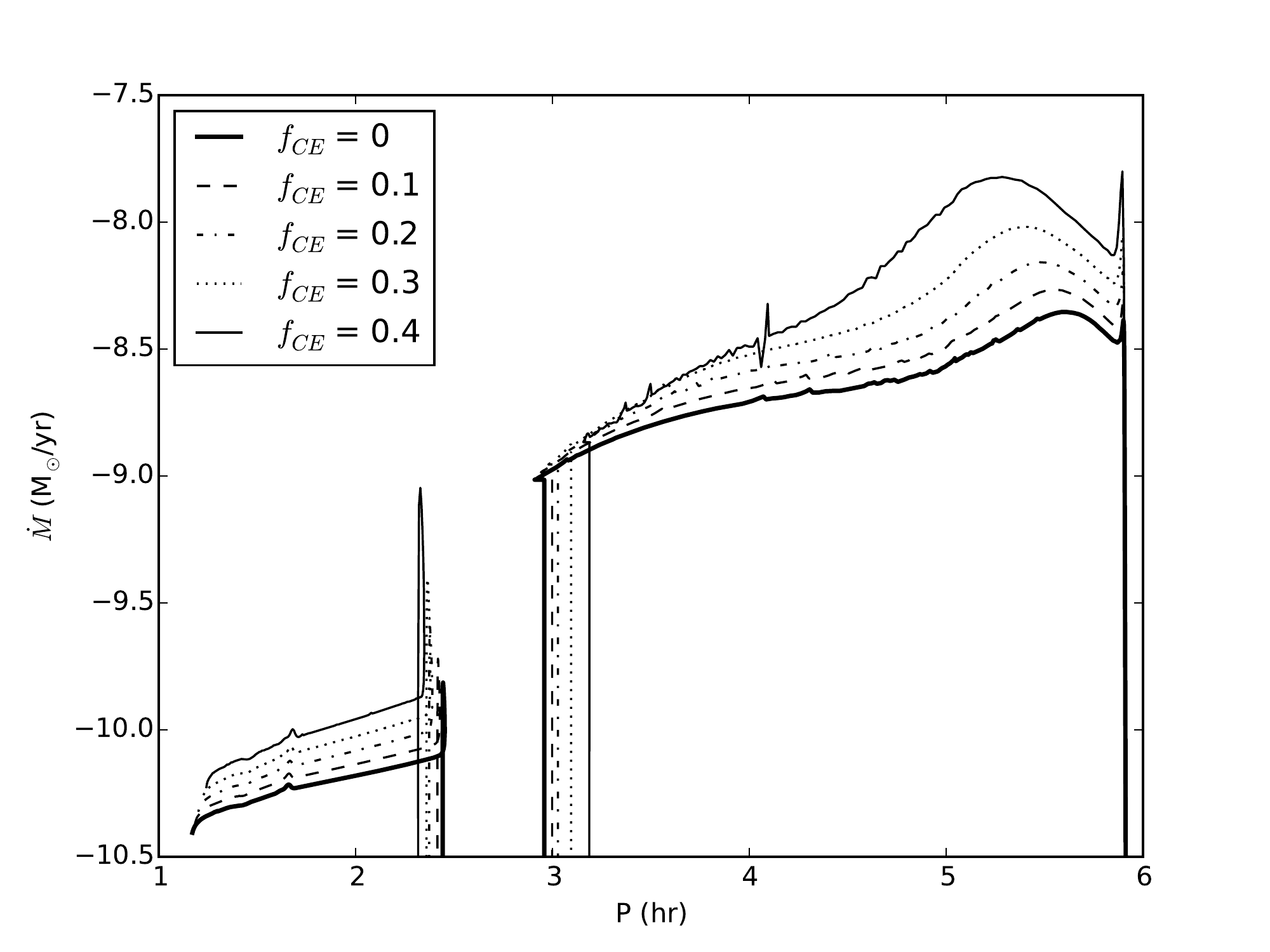}
\caption{Mass-transfer rate as function of periods for different
  values of $\delta$. The system initially consists of a 1.0 \msun WD
  with a 0.8 \msun MS companion.}
\label{fig:P_Mdot}
\end{figure}

Observationally, the effect of both the common-envelope like ejection
as well as (in most cases) the asymmetric mass loss would be an
enhancement of the mass-transfer rate and mass loss from the
system. To show this, we plot in Fig.~\ref{fig:P_Mdot} the period --
mass-transfer rate evolution of a system that initially consists of a
1.0 \msun WD and an 0.8 \msun donor, for different values of $f_{\rm
  CE}$. The mass transfer rate increases significantly, although we
have to caution that in these calculations the standard magnetic
braking laws are used that likely overestimate the mass-transfer rate
\citep[see][and references therein]{2011ApJS..194...28K}.  For the
eccentric system, the strong orbital modulation of the mass-transfer
rate is likely severely damped by the accretion disk, which provides a
buffer between the instantaneous mass-transfer rate and the brightness
of the system. \citet{2013MNRAS.434.1902P} make an interesting case
for the CV BK Lyn to be a system where, following a nova outburst 2000
years ago, the system has had a long phase of much higher
mass-transfer rate, only now coming down into the regime of low-mass
transfer dwarf novae. The ER UMa class of CVs in that picture would be
slightly older ``post-novae''. They also suggest that the finding of
\citet{2010AJ....139.1831S} that some systems are significantly
brighter after a nova outburst, while others not, is due to the same
effect and that this occurs only in short-period systems.

A second observational effect would be a change in the orbital period
after a nova outburst \citep{1983ApJ...268..710S}. In case of the
common-envelope like ejection the period would decrease by a factor
that follows from eq.~(\ref{eq:J_CE}) and the relative change ($\Delta
P/P$) is roughly a factor 10 larger than the relative mass change
($\Delta M_{\rm ej}/M_{\rm tot}$), i.e. $f_{\rm CE} \times 10^{-3} -
10^{-4}$. For the asymmetric mass ejection, the period could both
increase and decrease, within a factor few from the relative mass
change. There are very few measurements of period changes, showing
both increases and decreases
\citep{1983ApJ...268..710S,2011ApJ...742..112S}, but future
determinations, in particular for different types of systems could be
used to measure the relative importance of mass and angular momentum
loss from the systems.

Another way to test our hypothesis, is whether there is any
observational signature that could be used to find the systems that
experience unstable mass transfer and thus merge. The merged product
would most likely form some kind of low-mass giant star, where the WD
becomes the core and the MS star formed the envelope. They would be
vastly outnumbered by ordinary giants. Perhaps if via asteroseismology
we could measure the core/envelope mass ratio, some of them would
stand out as having a very high ratio compared to ordinary giants
evolved from single stars.

Finally, we mention that the higher mass- and angular-momentum loss
needed to explain the lack of low-mass WDs in CVs also eases the
discrepancy between the theoretical and observed period minimum
\citep[see e.g.][]{2011ApJS..194...28K,2015arXiv151004294S}, because
higher mass-transfer rates lead to a period minimum at longer period,
as can be seen in Fig.~\ref{fig:P_Mdot}.

\section{Conclusions}\label{conclusion}

We study the mass-transfer stability of binary systems in which a MS
star starts mass transfer to a WD to become a CV. Motivated by the
problem that the WD masses in CVs are higher than in pre-CVs and that
their space density seems significantly lower than theoretically
predicted, we investigate whether the influence of nova outbursts on
the stability of the mass transfer could selectively remove the
pre-CVs with low-mass WDs so that only the systems with massive WDs
remain. Interaction between the expanding nova envelope and the
companion may lead to a common-envelope like phase that could take
away angular momentum. Low-mass WDs are more prone to this instability
and can be effectively removed from the CV population if some 40 per
cent of the ejection energy is provided by the orbital
interaction. However, more massive WD would also be affected and for
this mechanism to work comfortably, the higher ejecta velocities
expected and observed for more massive WD, should lead to less
interaction with the companion.

We also investigate the influence of any asymmetry of the mass
ejection in the nova and find that for low-mass WD this can
significantly influence the orbit. The induced a small eccentricity
drives up the average mass-transfer rate, maybe even to a regime where
the material burns directly on the WD when it arrives, as a super-soft
X-ray source. However, it depends strongly on the magnitude of the
asymmetry and we find that the tidal circularization time scale in our
simplified models, is always significantly shorter than the time to
the next nova outburst, but it may explain the temporary mass-transfer
rate increase inferred by \citet{2013MNRAS.434.1902P} for BK Lyn and
the ER UMa systems. A more detailed and systematic investigation of
asymmetric mass loss in CVs is needed to assess its potential
influence on the CV population.

We conclude that is seems possible that the pre-CVs with low-mass WDs
do not make it to become CVs, because the first (few) nova outburst(s)
drive additional angular momentum loss that leads to unstable mass
transfer and merger of the system. As also suggested by
\citet{2015arXiv151004294S}, this would significantly decrease the
total space density of CVs and may make the theoretical WD mass
distribution in CVs consistent with the observations.

\section*{Acknowledgements}

We thank Ken Shen as well as Matthias Schreiber and the other
participants of the New York CV workshop for inspiration and helpful
discussions. ESP and GN thank the Radboud Excellence Initiative for
supporting ESP's stay at Radboud University. LS is Senior FNRS
associate.

\bibliographystyle{apj}
\bibliography{journals,binaries} 

\begin{thebibliography}{47}
\expandafter\ifx\csname natexlab\endcsname\relax\def\natexlab#1{#1}\fi

\bibitem[{Brandt \& Podsiadlowski(1995)}]{bp95}
Brandt, N. \& Podsiadlowski, P. 1995, MNRAS, 274, 461

\bibitem[{{Chesneau} \& {Banerjee}(2012)}]{2012BASI...40..267C}
{Chesneau}, O. \& {Banerjee}, D.~P.~K. 2012, Bulletin of the Astronomical
  Society of India, 40, 267

\bibitem[{{Chomiuk} {et~al.}(2014){Chomiuk}, {Linford}, {Yang}, {O'Brien},
  {Paragi}, {Mioduszewski}, {Beswick}, {Cheung}, {Mukai}, {Nelson}, {Ribeiro},
  {Rupen}, {Sokoloski}, {Weston}, {Zheng}, {Bode}, {Eyres}, {Roy}, \&
  {Taylor}}]{2014Natur.514..339C}
{Chomiuk}, L., {Linford}, J.~D., {Yang}, J., {et~al.} 2014, \nat, 514, 339

\bibitem[{{Davis} {et~al.}(2013){Davis}, {Siess}, \&
  {Deschamps}}]{2013A&A...556A...4D}
{Davis}, P.~J., {Siess}, L., \& {Deschamps}, R. 2013, \aap, 556, A4

\bibitem[{de~Kool(1992)}]{dek92}
de~Kool, M. 1992, \aap, 261, 188

\bibitem[{Eggleton(1983)}]{egg83}
Eggleton, P.~P. 1983, \apj, 268, 368

\bibitem[{{G{\"a}nsicke} {et~al.}(2009){G{\"a}nsicke}, {Dillon}, {Southworth},
  {Thorstensen}, {Rodr{\'{\i}}guez-Gil}, {Aungwerojwit}, {Marsh}, {Szkody},
  {Barros}, {Casares}, {de Martino}, {Groot}, {Hakala}, {Kolb}, {Littlefair},
  {Mart{\'{\i}}nez-Pais}, {Nelemans}, \& {Schreiber}}]{2009MNRAS.397.2170G}
{G{\"a}nsicke}, B.~T., {Dillon}, M., {Southworth}, J., {et~al.} 2009, \mnras,
  397, 2170

\bibitem[{{Kalogera}(1996)}]{1996ApJ...471..352K}
{Kalogera}, V. 1996, \apj, 471, 352

\bibitem[{{Kato} \& {Hachisu}(2009)}]{2009ApJ...699.1293K}
{Kato}, M. \& {Hachisu}, I. 2009, \apj, 699, 1293

\bibitem[{{Kato} \& {Hachisu}(2011)}]{2011ApJ...743..157K}
---. 2011, \apj, 743, 157

\bibitem[{{Knigge} {et~al.}(2011){Knigge}, {Baraffe}, \&
  {Patterson}}]{2011ApJS..194...28K}
{Knigge}, C., {Baraffe}, I., \& {Patterson}, J. 2011, \apjs, 194, 28

\bibitem[{Kolb(1993)}]{kol93}
Kolb, U. 1993, \aap, 271, 149

\bibitem[{{Livio}(1992)}]{1992ASPC...22..316L}
{Livio}, M. 1992, in Astronomical Society of the Pacific Conference Series,
  Vol.~22, Nonisotropic and Variable Outflows from Stars, ed. L.~{Drissen},
  C.~{Leitherer}, \& A.~{Nota}, 316

\bibitem[{{Livio} {et~al.}(1991){Livio}, {Govarie}, \&
  {Ritter}}]{1991A&A...246...84L}
{Livio}, M., {Govarie}, A., \& {Ritter}, H. 1991, \aap, 246, 84

\bibitem[{Paczy\'nski(1976)}]{pac76}
Paczy\'nski, B. 1976, in Structure and Evolution of Close Binary Systems, ed.
  P.~Eggleton, S.~Mitton, \& J.~Whelan (Dordrecht: Kluwer), 75

\bibitem[{{Patterson} {et~al.}(2013){Patterson}, {Uthas}, {Kemp}, {de Miguel},
  {Krajci}, {Foote}, {Hambsch}, {Campbell}, {Roberts}, {Cejudo}, {Dvorak},
  {Vanmunster}, {Koff}, {Skillman}, {Harvey}, {Martin}, {Rock}, {Boyd},
  {Oksanen}, {Morelle}, {Ulowetz}, {Kroes}, {Sabo}, \&
  {Jensen}}]{2013MNRAS.434.1902P}
{Patterson}, J., {Uthas}, H., {Kemp}, J., {et~al.} 2013, \mnras, 434, 1902

\bibitem[{{Paxton} {et~al.}(2013){Paxton}, {Cantiello}, {Arras}, {Bildsten},
  {Brown}, {Dotter}, {Mankovich}, {Montgomery}, {Stello}, {Timmes}, \&
  {Townsend}}]{2013ApJS..208....4P}
{Paxton}, B., {Cantiello}, M., {Arras}, P., {et~al.} 2013, \apjs, 208, 4

\bibitem[{{Paxton} {et~al.}(2015){Paxton}, {Marchant}, {Schwab}, {Bauer},
  {Bildsten}, {Cantiello}, {Dessart}, {Farmer}, {Hu}, {Langer}, {Townsend},
  {Townsley}, \& {Timmes}}]{2015ApJS..220...15P}
{Paxton}, B., {Marchant}, P., {Schwab}, J., {et~al.} 2015, \apjs, 220, 15

\bibitem[{{Politano}(1996)}]{1996ApJ...465..338P}
{Politano}, M. 1996, \apj, 465, 338

\bibitem[{Politano \& Webbink(1989)}]{pw89}
Politano, M. \& Webbink, R.~F. 1989, in IAU Colloq., Vol. 114, White Dwarfs,
  440--442

\bibitem[{{Pretorius}(2014)}]{2014xru..confE.164P}
{Pretorius}, M. 2014, in The X-ray Universe 2014, 164

\bibitem[{Rappaport {et~al.}(1983)Rappaport, Verbunt, \& Joss}]{rvj83}
Rappaport, S., Verbunt, F., \& Joss, P.~C. 1983, \apj, 275, 713

\bibitem[{{Repetto} \& {Nelemans}(2015)}]{2015MNRAS.453.3341R}
{Repetto}, S. \& {Nelemans}, G. 2015, \mnras, 453, 3341

\bibitem[{{Ribeiro} {et~al.}(2013){Ribeiro}, {Munari}, \&
  {Valisa}}]{2013ApJ...768...49R}
{Ribeiro}, V.~A.~R.~M., {Munari}, U., \& {Valisa}, P. 2013, \apj, 768, 49

\bibitem[{{Ritter}(1988)}]{1988A&A...202...93R}
{Ritter}, H. 1988, \aap, 202, 93

\bibitem[{{Schaefer}(2011)}]{2011ApJ...742..112S}
{Schaefer}, B.~E. 2011, \apj, 742, 112

\bibitem[{{Schaefer} \& {Collazzi}(2010)}]{2010AJ....139.1831S}
{Schaefer}, B.~E. \& {Collazzi}, A.~C. 2010, \aj, 139, 1831

\bibitem[{{Schaefer} \& {Patterson}(1983)}]{1983ApJ...268..710S}
{Schaefer}, B.~E. \& {Patterson}, J. 1983, \apj, 268, 710

\bibitem[{{Schenker} \& {King}(2002)}]{2002ASPC..261..242S}
{Schenker}, K. \& {King}, A.~R. 2002, in Astronomical Society of the Pacific
  Conference Series, Vol. 261, The Physics of Cataclysmic Variables and Related
  Objects, ed. B.~T. {G{\"a}nsicke}, K.~{Beuermann}, \& K.~{Reinsch}, 242

\bibitem[{{Schreiber} {et~al.}(2015){Schreiber}, {Zorotovic}, \&
  {Wijnen}}]{2015arXiv151004294S}
{Schreiber}, M.~R., {Zorotovic}, M., \& {Wijnen}, T.~P.~G. 2015, ArXiv e-prints

\bibitem[{{Shara} {et~al.}(1986){Shara}, {Livio}, {Moffat}, \&
  {Orio}}]{1986ApJ...311..163S}
{Shara}, M.~M., {Livio}, M., {Moffat}, A.~F.~J., \& {Orio}, M. 1986, \apj, 311,
  163

\bibitem[{{Shen}(2015)}]{2015ApJ...805L...6S}
{Shen}, K.~J. 2015, \apjl, 805, L6

\bibitem[{{Siess} {et~al.}(2013){Siess}, {Izzard}, {Davis}, \&
  {Deschamps}}]{2013A&A...550A.100S}
{Siess}, L., {Izzard}, R.~G., {Davis}, P.~J., \& {Deschamps}, R. 2013, \aap,
  550, A100

\bibitem[{Soberman {et~al.}(1997)Soberman, Phinney, \& van~den Heuvel}]{sph97}
Soberman, G.~E., Phinney, E.~S., \& van~den Heuvel, P.~J. 1997, \aap, 327, 620

\bibitem[{{Starrfield} {et~al.}(1972){Starrfield}, {Truran}, {Sparks}, \&
  {Kutter}}]{1972ApJ...176..169S}
{Starrfield}, S., {Truran}, J.~W., {Sparks}, W.~M., \& {Kutter}, G.~S. 1972,
  \apj, 176, 169

\bibitem[{{Tauris} \& {van den Heuvel}(2006)}]{2006csxs.book..623T}
{Tauris}, T.~M. \& {van den Heuvel}, E.~P.~J. 2006, {Formation and evolution of
  compact stellar X-ray sources}, ed. W.~H.~G. {Lewin} \& M.~{van der Klis},
  623--665

\bibitem[{{Toonen} \& {Nelemans}(2013)}]{2013A&A...557A..87T}
{Toonen}, S. \& {Nelemans}, G. 2013, \aap, 557, A87

\bibitem[{{Toonen} {et~al.}(2014){Toonen}, {Voss}, \&
  {Knigge}}]{2014MNRAS.441..354T}
{Toonen}, S., {Voss}, R., \& {Knigge}, C. 2014, \mnras, 441, 354

\bibitem[{{Townsley} \& {Bildsten}(2004)}]{tb04}
{Townsley}, D.~M. \& {Bildsten}, L. 2004, \apj, 600, 390

\bibitem[{{Townsley} \& {Bildsten}(2005)}]{2005ApJ...628..395T}
---. 2005, \apj, 628, 395

\bibitem[{van~den Heuvel {et~al.}(1992)van~den Heuvel, Bhattacharya, Nomoto, \&
  Rappaport}]{hbn+92}
van~den Heuvel, E. P.~J., Bhattacharya, D., Nomoto, K., \& Rappaport, S.~A.
  1992, \aap, 262, 97

\bibitem[{Verbunt \& Phinney(1995)}]{vp95}
Verbunt, F. \& Phinney, E.~S. 1995, \aap, 296, 709

\bibitem[{Warner(1995)}]{war95b}
Warner, B. 1995, "Cataclysmic variable stars" (Cambridge: CUP)

\bibitem[{{Wijnen} {et~al.}(2015){Wijnen}, {Zorotovic}, \&
  {Schreiber}}]{2015A&A...577A.143W}
{Wijnen}, T.~P.~G., {Zorotovic}, M., \& {Schreiber}, M.~R. 2015, \aap, 577,
  A143

\bibitem[{{Williams}(2013)}]{2013IAUS..281...88W}
{Williams}, R. 2013, in IAU Symposium, Vol. 281, IAU Symposium, ed. R.~{Di
  Stefano}, M.~{Orio}, \& M.~{Moe}, 88--95

\bibitem[{{Woudt} {et~al.}(2009){Woudt}, {Steeghs}, {Karovska}, {Warner},
  {Groot}, {Nelemans}, {Roelofs}, {Marsh}, {Nagayama}, {Smits}, \&
  {O'Brien}}]{2009ApJ...706..738W}
{Woudt}, P.~A., {Steeghs}, D., {Karovska}, M., {et~al.} 2009, \apj, 706, 738

\bibitem[{{Zorotovic} {et~al.}(2011){Zorotovic}, {Schreiber}, \&
  {G{\"a}nsicke}}]{2011A&A...536A..42Z}
{Zorotovic}, M., {Schreiber}, M.~R., \& {G{\"a}nsicke}, B.~T. 2011, \aap, 536,
  A42

\end{thebibliography}

\end{document}